\documentclass{aastex}
\usepackage{spr-astr-addons}
\usepackage{url}
\usepackage{lscape, morefloats, graphicx, float} 

\RequirePackage{color}

\begin{document}

\title{The Galactic kinematics of cataclysmic variables}
\slugcomment{Not to appear in Nonlearned J., 45.}
\shorttitle{The Galactic kinematics of CVs}
\shortauthors{T. Ak, S. Bilir, A. \"Ozd\"onmez, F. Soydugan, E. Soydugan, \c C. P\"usk\"ull\"u, S. Ak, \and Z. Eker}

\author{T. Ak \altaffilmark{1}}
\altaffiltext{1}{Istanbul University, Faculty of Science, Department 
of Astronomy and Space Sciences, 34119 University, Istanbul, Turkey\\
\email{tanselak@istanbul.edu.tr}}

\author{S. Bilir \altaffilmark{1}}
\altaffiltext{1}{Istanbul University, Faculty of Science, Department 
of Astronomy and Space Sciences, 34119 University, Istanbul, Turkey\\}
\and
\author{A. \"Ozd\"onmez\altaffilmark{2}} 
\altaffiltext{2}{Istanbul University, Graduate School of Science and 
Engineering, Department of Astronomy and Space Sciences, 34116 Beyaz\i t, 
Istanbul, Turkey\\}
\and
\author{F. Soydugan\altaffilmark{3}} 
\altaffiltext{3}{\c Canakkale Onsekiz Mart University, Faculty of 
Sciences and Arts, Department of Physics, 17100 \c Canakkale, Turkey\\}
\altaffiltext{4}{\c Canakkale Onsekiz Mart University, Astrophysics 
Research Center and Ulup\i nar Observatory, 17100 \c Canakkale, Turkey\\}
\and
\author{E. Soydugan\altaffilmark{3}} 
\altaffiltext{3}{\c Canakkale Onsekiz Mart University, Faculty of 
Sciences and Arts, Department of Physics, 17100 \c Canakkale, Turkey\\}
\altaffiltext{4}{\c Canakkale Onsekiz Mart University, Astrophysics 
Research Center and Ulup\i nar Observatory, 17100 \c Canakkale, Turkey\\}
\and
\author{\c C. P\"usk\"ull\"u\altaffilmark{3}} 
\altaffiltext{3}{\c Canakkale Onsekiz Mart University, Faculty of 
Sciences and Arts, Department of Physics, 17100 \c Canakkale, Turkey\\}
\altaffiltext{4}{\c Canakkale Onsekiz Mart University, Astrophysics 
Research Center and Ulup\i nar Observatory, 17100 \c Canakkale, Turkey\\}
\and
\author{S. Ak\altaffilmark{1}} 
\altaffiltext{1}{Istanbul University, Faculty of Science, Department 
of Astronomy and Space Sciences, 34119 University, Istanbul, Turkey\\}
\and
\author{Z. Eker\altaffilmark{4}} 
{\altaffiltext{4}{Akdeniz University, Faculty of Sciences, Space Science 
and Technologies Department, 07058 Campus, Antalya, Turkey}

\begin{abstract}
Kinematical properties of CVs were investigated according to population types 
and orbital periods, using the space velocities computed from recently updated 
systemic velocities, proper motions and parallaxes. Reliability of collected 
space velocity data are refined by removing 34 systems with largest space 
velocity errors. The 216 CVs in the refined sample were shown to have 
a dispersion of 53.70 $\pm$ 7.41 km s$^{-1}$ corresponding to a mean 
kinematical age of 5.29 $\pm$ 1.35 Gyr. Population types of CVs were identified 
using their Galactic orbital parameters. According to the population analysis, 
seven old thin disc, nine thick disc and one halo CV were found in the sample, 
indicating that 94\% of CVs in the Solar Neighbourhood belong to the 
thin-disc component of the Galaxy. Mean kinematical ages $3.40 \pm 1.03$ 
and $3.90 \pm 1.28$ Gyr are for the non-magnetic thin-disc CVs below and 
above the period gap, respectively. There is not a meaningful difference between 
the  velocity dispersions below and above the gap. Velocity dispersions 
of the non-magnetic thin-disc systems below and above the gap are 
$24.95 \pm 3.46$ and $26.60 \pm 4.18$ km s$^{-1}$, respectively. This 
result is not in agreement with the standard formation and evolution theory of CVs. 
The mean kinematical ages of the CV groups in various orbital period intervals 
increase towards shorter orbital periods. This is in agreement with the 
standard theory for the evolution of CVs. Rate of orbital period change was 
found to be $dP/dt=-1.62(\pm 0.15)\times 10^{-5}$ sec yr$^{-1}$. 
\end{abstract}


\section{Introduction}

A cataclysmic variable (hereafter CV) consists of a white dwarf primary and 
a low-mass secondary which overflows its Roche lobe. Material from the donor star 
is transferred to the primary usually via a gas stream and an accretion disc. 
The white dwarf in a magnetic CV accretes material through accretion channels and 
columns instead of an accretion disc formation of which is prevented by the strong 
magnetic field of the primary component in the system.

Standard evolution theory of CVs proposes that a CV begins its evolution as a detached 
main-sequence binary star with a more massive primary.  
Nuclear evolution of the primary component drives the system to a common envelope 
(CE) phase \citep{KS96} during which the envelope of the giant star is ejected as 
the dynamical friction extracts orbital angular momentum. Evolution of the 
binary system to the shorter orbital periods after the CE phase is thought to be governed by 
orbital angular momentum loses due to the gravitational radiation \citep{Pac67} and 
the magnetic braking \citep{VZ81,Retal82,Retal83,PS83,SR83,K88} through the post-CE 
and CV phases. Dynamical evolution of CVs can be studied using their orbital 
period distribution since the orbital period ($P$) is the most precisely determined 
orbital parameter for these systems. The most striking features of the CV period 
distribution are the orbital period gap between roughly 2 and 3 h 
\citep{SR83,K88,Kn11} and a sharp cut-off at about 80 min, period minimum 
\citep{Hametal1988,Wiletal05,Ganetal09}. 

Although predictions of the population synthesis models based on the 
standard CV evolution theory can be tested using data sets obtained from photometric 
observations \citep[see][and references therein]{Aketal08,Aketal10,Ozdonetal15}, 
these data sets are strongly biased by the selection effects, primarily 
the brightness dependent ones \citep{Pretetal07}. Nevertheless, the 
age distribution of CVs is not biased by brightness-selection 
\citep{Kol01}, since the age of a CV does not affect its mass transfer 
rate at a given orbital period. Thus, the kinematical ages of CV groups 
can be used to test the predictions of the model. The kinematical 
age is defined as the time span since formation of component stars.

From the age structure of a Galactic CV population obtained 
by applying standard models for the formation and evolution of these 
systems \citep{KS96}, it is predicted that CVs above the orbital period 
gap ($P\gtrsim$ 3~h) must have an average age of 1~Gyr, while the mean 
age of systems below the gap ($P\lesssim$ 2~h) should be 3-4~Gyr 
\citep[see also][]{RB86}. This age difference is mainly due to the time 
spent evolving from the post-CE phase into contact \citep{Kol01}. 
\cite{KS96} predicted using a relation between age ($t$) and total space velocity 
dispersion ($\sigma_{\nu}$) of field stars that the dispersions of the systemic radial 
velocities ($\gamma$) for the systems above and below the orbital period gap 
are $\sigma_{\gamma} \simeq 15$ and $\sigma_{\gamma} \simeq 30$ 
km~s$^{-1}$, respectively. 

The observational tests of the predictions of the population studies 
based on standard formation and evolution model of CVs can be done by 
estimating kinematical properties of the systems. The predicted difference 
between the velocity dispersions for the systems above and below the 
period gap could not be detected by \cite{vPaetal96}, 
who analysed the observed $\gamma$ velocities for a sample of CVs. 
\cite{Aketal10} found a dispersion $\sigma_{\gamma} = 30\pm 5$ km~s$^{-1}$ 
for the systems below the period gap, a value proper to the predictions. 
However, they could not detect a considerable dispersion difference between 
the systems above and below the period gap, as they calculated a dispersion 
of  $\sigma_{\gamma} = 26\pm 4$ km~s$^{-1}$ for CVs above the gap. 
Interestingly, according to \cite{Kol01} if magnetic braking does not operate 
in the detached phase, the $\gamma$ velocity dispersions should be 
$\sigma_{\gamma} \simeq 27$ and $\sigma_{\gamma} \simeq 32$ km~s$^{-1}$ for 
the systems above and below the gap, respectively. Considering the kinematical 
ages determined by \cite{Aketal10} for non-magnetic CVs below and above the 
period gap, which are $5\pm 1$ and $4\pm 1$ Gyr, respectively, it is  clear 
that the difference between these ages is not as large as expected from the 
standard evolution theory of CVs. Note that \cite{Peters08} estimated mean 
kinematical ages of $\geq$5, $\geq$4 and $\geq$6 Gyr for all CVs, non-magnetic 
and magnetic systems in a large sample, respectively. \cite{Peters07,Peters08} 
concludes that kinematics of all CVs in his sample are indicative of 
a moderately old thin-disc Galactic population. A similar result was found by 
\cite{Aketal13} who concluded from the Galactic orbital parameters of 159 CVs 
in the Solar Neighbourhood that 94$\%$ of CVs are thin-disc members and the 
rest are thick-disc stars. 

Kinematical properties for a Galactic population of CVs can be only found having reliable 
distance, astrometry and systemic velocity measurements. Although there are 
methods for determining the population types based on probability distributions 
\citep[i.e.][]{Bensetal03,Bensetal05}, reliable population types can be 
determined from Galactic orbits of the objects \citep{Aketal13}. \cite{Aketal10} 
determined kinematical properties of CVs with systemic velocities collected 
from the literature and distances estimated from PLCs (Period-Luminosity-Colours) 
relation \citep{Aketal07a,Aketal08}. They determined population types of CVs in 
their sample using probability distributions described by 
\cite{Bensetal03,Bensetal05} instead of Galactic orbits of the systems.

Number of CVs with measured systemic velocities increased from 194 to 250 since 
the study of \cite{Aketal10}. In addition, a new PLCs relation has been suggested 
by \cite{Ozdonetal15}. They used both the Two Micron All Sky Survey 
\citep[2MASS;][]{Sktretal2006} and Wide-field Infrared Survey Explorer 
\citep[$WISE$;][]{Wrietal2010} photometry to predict the distances of CVs. 
With this new PLCs relation, absolute magnitudes of CVs can be estimated 
$\sim$2 times more precise as compared to the PLCs relation of \cite{Aketal08}. 
These new $\gamma$ velocity data and availability of a better absolute 
magnitude, thus distance, prediction method motivated us to re-investigate the 
kinematical properties of CVs in terms of the Galactic populations and orbital 
period. We determined population types of CVs using a pure dynamical 
approximation. Thus, in this paper we aim to derive kinematical age profiles, 
space velocity dispersions and $\gamma$ velocity dispersions of CV groups 
according to different orbital period regimes and the Galactic populations 
in order to test the predictions of the population models and to understand 
orbital period evolution of CVs. 

\section{The data}

The distances (or parallaxes), proper motions and systemic velocities 
are basic parameters to compute the space velocity of a star. The CV sample, which 
are already collected by  \cite{Ozdonetal15}, was preferred as a homogeneous sample 
of CVs regarding to the distances. Then, a new additional list was constructed by 
searching and collecting new and more systems with proper motions and systemic 
velocities ($\gamma$) from the literature.

\subsection{Systemic velocities}

Since orbits of CVs are circular, it is conventional to use 
$V_{r}(\phi)=\gamma+K_{1,2}\sin\phi$ to express instantaneous radial velocity of 
a component. Here, $\gamma$ is the center of mass radial velocity of a CV. In this 
equation, $\phi$  is the orbital phase, $K_{1,2}$ are the semi-amplitudes of the radial 
velocity variation, where 1 and 2 are primary and secondary components, respectively.

\cite{Aketal10} collected $\gamma$ velocities of CVs published in the literature 
up to the middle of the year 2007 and combined their $\gamma$ velocity collection 
with the sample of \cite{vPaetal96} who collected systemic velocities of CVs from 
the literature covering times up to the year 1994. In this study, $\gamma$ velocities 
up to the middle of the year 2014 were collected in a similar manner. The same 
criteria defined by \cite{vPaetal96} and \cite{Aketal10} were adopted: (1) If there 
are more than one determination of $\gamma$ velocity for a system, an average of 
$\gamma$ velocities is taken, (2) a new average value is calculated if there is 
a new measurement which is not included in the previous lists, (3) if there are more 
than one $\gamma$ velocity measurement from different methods in a study, then the 
value recommended by the author was taken, (4) since very large variations in radial 
velocities can be observed during superoutbursts, $\gamma$ velocities obtained during 
superoutbursts of SU UMa type dwarf novae have been ignored. We have listed 250 
CVs with known parallax, proper motions and systemic velocity in Table 1.

The radial velocities from the absorption lines of secondary component represent
the system best, because emission lines originate mostly in the accretion
disc. The radial velocities derived from emission lines are likely affected by the
motions in the accretion disc or the matter stream. In addition, $\gamma$ velocities 
of magnetic CVs can be affected by infalling material along magnetic field lines
\citep{Peters07,Peters08}. Therefore, the  velocities derived from emission lines
($\gamma_{em}$) may not be reliable \citep{Noretal02}. Consequently, the  velocities
measured from absorption lines ($\gamma_{abs}$) for 80 systems are the most reliable 
values used in the analyses.

It is clear that results of the kinematical analyses could be biased due to 
possible systematic errors in the systemic velocities coming from emission lines. 
Thus, systematic and statistical accuracy of the $\gamma_{em}$ values must be 
studied. In the sample there are 53 systems with both the 
$\gamma_{em}$ and $\gamma_{abs}$ measurements. Average of the difference of these 
values is $<\gamma_{em}-\gamma_{abs}>=-3.5\pm 22.6$ km s$^{-1}$, where the 
error is the standard deviation of the distribution of individual differences. 
It should be noted that the median value of $(\gamma_{em}-\gamma_{abs})$ is 
-3.6 km s$^{-1}$. There were only 10 systems for \cite{vPaetal96} 
to estimate $<\gamma_{em}-\gamma_{abs}>=+2.5\pm 13.8$ km s$^{-1}$. They 
concluded that there is no considerable systematic difference between the 
systemic velocities derived from the emission and the absorption lines. From the 
average difference found in this study, we too conclude that meaningful statistical 
analyses can be done. Error histogram for the $\gamma$ velocities is shown in 
Fig. 1a. The median value and standard deviation of $\gamma$ velocity errors 
are 5.00 and $\pm$6.62 km~s$^{-1}$, respectively. 

\begin{landscape}  
\textwidth = 700 pt
\begin{table*}
\setlength{\tabcolsep}{.5pt} 
\begin{center}
\tiny{
\caption{Names, coordinates ($\alpha_{2000.0}$, $\delta_{2000.0}$), types, orbital 
periods ($P$), parallaxes ($\pi$), proper motions ($\mu_{\alpha}\cos \delta,\mu_{\delta}$) 
and systemic velocities ($\gamma$) of CVs in the preliminary sample. In column Type-1, CV 
denotes CVs with unknown types, DN dwarf novae, NL nova-like stars and N novae. Column 
Type-2 indicates magnetic (M, polars and intermediate polars) and non-magnetic systems 
(nM). The last column is for references. The first number in the last column is for 
the parallax, the second for the proper motion. The bibliographic codes are for 
the $\gamma$ velocities. Full table can be obtained electronically.}
\begin{tabular}{lccccccccccccccl}
\hline
ID & Name & $\alpha_{2000.0}$ & $\delta_{2000.0}$        & Type-1  & Type-2  & $P$ & $\pi$  & err  & $\mu_{\alpha}\cos \delta$ & err     & $\mu_{\delta}$ & err     & $\gamma$  & err&  \multicolumn{1}{c}{References}   \\
   &      &    (hh~mm~ss)     & ($^{\circ}$~~~$'$~~~$''$)  &         &         & (h) & (mas)  & (mas)&     (mas~yr$^{-1}$)              & (mas~yr$^{-1}$)&   (mas~yr$^{-1}$)     & (mas~yr$^{-1}$)& (km s$^{-1}$) & (km s$^{-1}$) &     \\
\hline
  1 &     WW Cet & 00:11:24.78 & -11:28:43.10 & DN & nM & 0.1758   &  4.03 &  0.54 &   15.0 & 6.1 &    11.6 & 1.5 &  22.9 &  6.2 & (1, 3, 1996AJ....111.2077R; 1996A\&A...312...93V; 1997A\&A...327..231T) \\
  2 &   V592 Cas & 00:20:52.22 & +55:42:16.30 & NL & nM & 0.115063 &  3.34 &  0.34 &  -13.7 & 4.0 &    -3.9 & 4.1 &  21.0 & 14.0 & (2, 4, 1998PASP..110..784H) \\
  3 &   V709 Cas & 00:28:48.83 & +59:17:22.00 & NL & M  & 0.222204 &  3.11 &  0.42 &    0.2 & 2.7 &    -1.8 & 1.9 & -41.0 &  3.0 & (1, 3, 2010PASP..122.1285T) \\
  4 &     PX And & 00:30:05.81 & +26:17:26.40 & NL & nM & 0.146353 &  1.24 &  0.17 &   -7.5 & 2.7 &    -9.6 & 3.0 & -18.3 & 24.8 & (1, 3, 1995MNRAS.273..863S; 1996A\&A...312...93V) \\
  5 &    LTT 560 & 00:59:28.89 & -26:31:05.50 & CV & nM & 0.1475   &  9.06 &  1.22 &  136.3 & 4.0 &  -246.4 & 4.0 & 36.51 & 0.72 & (1, 5, 2011A\&A...532A.129T) \\
 ... &        ...&         ... &          ... &     ... &      ... &   ... &   ... &    ... & ... &     ... & ... &   ... &  ... &        ... &   ... \\
 ... &        ...&         ... &          ... &     ... &      ... &   ... &   ... &    ... & ... &     ... & ... &   ... &  ... &        ... &   ... \\
 ... &        ...&         ... &          ... &     ... &      ... &   ... &   ... &    ... & ... &     ... & ... &   ... &  ... &        ... &   ... \\
\hline     
\end{tabular}
(1) 2015NewA...34..234O, (2) 2007NewA...12..446A, (3) 2013AJ....145...44Z, (4) 2000A\&A...355L..27H, (5) 2010AJ....139.2440R}
\end{center}
\end{table*}
\end{landscape}  

\begin{figure}[h]
\begin{center}
\includegraphics[scale=0.5, angle=0]{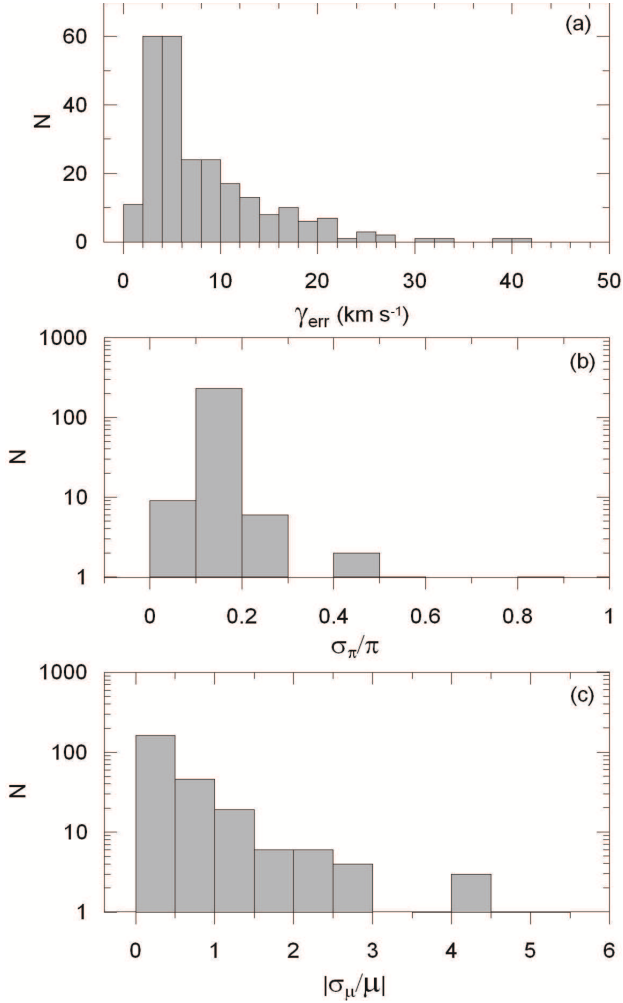}
\caption[] {\small Distribution of $\gamma$ velocity errors (a), relative parallax 
errors (b) and relative proper motion errors (c) of the present CV sample.}
\end{center}
\end{figure}

\subsection{Distances and proper motions}

Although precise trigonometric parallaxes of some CVs were measured by many authors 
\citep{Due99,McArtetal99,McArtetal01,Thor03,Beuetal2003,Beuetal2004,Harretal04,Roetal07,Thor08,Thor09}, 
the measured number of parallaxes is only about 30. Therefore, parallaxes of CVs were 
collected in the following way as a general rule. Trigonometric parallaxes were taken 
as they are, where available. A new PLCs relation, which uses $J$, $K_s$ and $W$1 
magnitudes in the Two Micron All Sky Survey \citep[2MASS;][]{Sktretal2006} and 
Wide-field Infrared Survey Explorer \citep[$WISE$;][]{Wrietal2010} photometry and 
the orbital periods, was established by \cite{Ozdonetal15} in order to estimate the 
distances of 313 CVs. This new PLCs relation, which is valid in the ranges 
$1.37\leq P$(h) $\leq 12$, $0.13 \leq (J-K_s)_{0} \leq 1.01$, 
$-0.36 \leq (K_{s}-W1)_{0} \leq 0.82$ and $2.9 < M_{J} < 10.4$ mag, was used to estimate 
the distances for CVs for which trigonometric parallaxes are not available. Here the 
subscript ``0'' denotes de-reddened colours. If a CV is out of the validity limits, then, 
the previous PLCs relation by \cite{Aketal07a}, involving the 2MASS photometry, was used. 
For detailed descriptions of the methods see \cite{Aketal07a,Aketal08} and 
\cite{Ozdonetal15}. Near and middle infrared magnitudes were taken from 
\cite{Cutetal2003,Cutetal2012}. Parallaxes of CVs were calculated using their distances 
with the well-known formula $\pi(mas) = 1000/d($pc$)$. CVs with orbital periods ($P$) 
longer than 12 h were discarded since a CV with orbital period longer than this limit 
possibly contains a secondary star on its way to becoming a red giant \citep{Hell2001}. 
Systems with $P < 80$ min were not included in our preliminary sample, as well, because 
these systems must contain a degenerate secondary star.  

The proper motions of the CVs in this study were taken from the {\it UCAC4} Catalogue of \cite{Zachetal13}, the 
{\it PPMXL} Catalogue of \cite{Roetal10}, the {\it Tycho-2} Catalogue of \cite{Hoetal00}, 
and from the re-reduced $Hipparcos$ catalogue of \cite{vLee07}. Fig. 1b-c show 
the distribution of relative parallax errors and relative proper motion errors, respectively. 
The median value and standard deviation of relative parallax errors are 0.14 and $\pm$0.11, 
respectively. The median value and standard deviation of proper motion 
errors are 0.34 and $\pm$0.94 mas, respectively. Parallaxes and proper motion 
components are listed in Table 1 together with observational uncertainties. 
The columns of the table are organized as name, equatorial coordinates, type of the 
CV, orbital period, parallax, proper motion components, and $\gamma$ velocity. The 
types, equatorial coordinates and orbital periods of CVs in Table 1 were taken 
from \citet[][Edition 7.7]{RK03}. 

\subsection{Galactic space velocities}

The algorithms and transformation matrices of \cite{JS87} were used to compute the space 
velocities with respect to the Sun. Equatorial 
coordinates ($\alpha$, $\delta$), proper motion components ($\mu_{\alpha}\cos\delta$, 
$\mu_{\delta}$), systemic velocity ($\gamma$) and the parallax ($\pi$) are the basic input 
data required. The form of this 
input data is adopted for the epoch of J2000 as described in the International Celestial 
Reference System (ICRS) of the $Hipparcos$ and the $Tycho$ Catalogues \citep{ESA1997}. 
The transformation matrices of \cite{JS87} use the notation of the right handed system. 
Therefore, the $U$, $V$ and $W$ are the components of a velocity 
vector of a star with respect to the Sun, where the $U$ is directed toward the Galactic 
Center ($l=0^{o}, b=0^{o}$), the $V$ is in the direction of the Galactic rotation 
($l=90^{o}, b=0^{o}$), and the $W$ is towards the North Galactic Pole ($b=90^{o}$). 
Here, $l$ and $b$ are the Galactic longitude and latitude, respectively. Although CVs 
in the sample are relatively close objects, that is in the Solar neighbourhood, corrections 
for differential Galactic rotation were applied to the space velocities as described 
in \cite{MB81}. Galactic space velocity components were also corrected for the Local 
Standard of Rest (LSR) by adding the space velocity of the Sun to the space velocity 
components of CVs. The adopted space velocity of the Sun is 
$(U,V,W)_{\odot}=(8.50,13.38,6.49)$ km~s$^{-1}$ \citep{Cosketal2011}. 

The uncertainties of the space velocity components were computed by propagating the 
uncertainties of the input data with the algorithm given by \cite{JS87}. The uncertainties 
of the total space velocities ($S_{err}=(U_{err}^{2}+V_{err}^{2}+W_{err}^{2})^{1/2}$) were 
also calculated. The histograms of the propagated uncertainties of total space velocities 
($S_{err}$) and their components ($U_{err}, V_{err}, W_{err}$) are displayed in Fig. 2 
with unshaded areas. The median and standard deviation of the total space velocity 
uncertainties in Fig. 2a are 15 and $\pm$15 km~s$^{-1}$, respectively. 

\begin{figure}[h]
\begin{center}
\includegraphics[scale=0.40, angle=0]{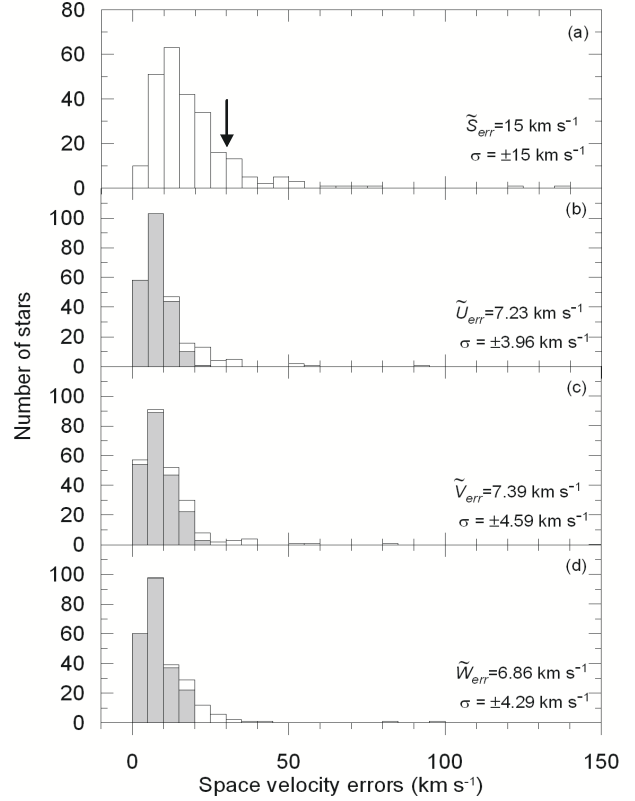}
\caption[] {\small The histograms of the propagated uncertainties of total space velocities 
($S_{err}$), and their components ($U_{err}, V_{err}, W_{err}$). Unshaded areas are for 
the preliminary sample, shaded areas for the refined sample.}
\end{center}
\end{figure}

As the space velocity dispersions and kinematical ages can be biased by the space velocities 
with very large uncertainties, we decided to remove all CVs with 
$S_{err} > 30$ km s$^{-1}$, which corresponds to the median plus standard deviation of the 
space velocity uncertainties, in order to refine the CV sample in this study. This limit is 
indicated with an arrow in Fig. 2a. CVs with errors above this limit were discarded, 
thus 216 CVs were left in the refined sample. Shaded areas in Fig. 2b-d display the 
histograms of the uncertainties of the space velocity components ($U_{err}, V_{err}, W_{err}$) 
calculated for CVs in the refined sample. The mean values and dispersions of space velocity 
components calculated for the CV groups are given in Table 3. The median values of 
the errors for the refined sample are $\widetilde U_{err} = 7.23$, 
$\widetilde V_{err} = 7.39$ and $\widetilde W_{err} = 6.86$ km s$^{-1}$ while error 
distributions have standard deviations $\pm 3.96$, $\pm 4.59$ and $\pm 4.29$ km s$^{-1}$, 
respectively for the $U$, $V$ and $W$ components of the space velocity. 

\subsection{Population analysis} 

In order to investigate kinematical and dynamical properties of CVs in the thin disc, 
thick disc and halo components of the Galaxy, a precise population analysis must be 
done. Here we adopt a pure dynamical approach to find population types of CVs in the 
refined sample. This approach is based on Galactic orbits of CVs. A similar method 
described in \cite{Aketal13} was used. Therefore, we first perform test-particle integration 
in a Milky Way potential which consists of a logarithmic halo of the form

\begin{eqnarray}
  \Phi_{\rm halo}(r)=v_{0}^{2} \ln \left(1+\frac{r^2}{d^2}\right).
\end{eqnarray}
Here, $v_{0}=186$ km s$^{-1}$ and $d=12$ kpc. A Miyamoto-Nagai potential represents the 
disc:

\begin{eqnarray}
  \Phi_{\rm disc}(R,z)=-\frac{G M_{\rm d}} { \sqrt{R^{2} + \left(
        a_d + \sqrt{z^{2}+b_d^{2}} \right)^{2}}},
\end{eqnarray}
with $M_{\rm d}=10^{11}~M_{\odot}$, $a_d=6.5$ kpc and $b_d=0.26$
kpc. Finally, the bulge is modelled as a Hernquist potential,

\begin{eqnarray}
  \Phi_{\rm bulge}(r)=-\frac{G M_{\rm b}} {r+c},
\end{eqnarray}
using $M_{\rm b}=3.4\times10^{10}~M_{\odot}$ and $c=0.7$ kpc. A good representation 
of the Milky Way is obtained with the superposition of these components. The orbital period of the LSR is 
taken $P=2.18\times10^8$ years while $V_c=222.5$ km s$^{-1}$ represents the circular 
rotational velocity at the Solar Galactocentric distance, $R_0=8$ kpc 
\citep{Coskunogluetal12,Bil12}. 

The population types of CVs were determined according to the maximum vertical distances 
from the Galactic plane ($Z_{max}$) for their calculated orbits within the integration time 
of 3 Gyr, i.e. backwards in time. This integration time 
corresponds to 12-15 revolutions around the Galactic center so that the averaged 
orbital parameters can be determined reliably. Although 
the thick disc component of the Galaxy was discovered more than 30 years ago 
\citep{GilReid1983}, there is still not a consensus for the numerical values of the 
parameters of this component. Especially, there is a degeneracy between the space density 
in the Solar Neighbourhood and the scale height of the thick disc component 
\citep{Siegetal2002,Karaalietal2004,Biletal2006,Biletal2008}. Thus, we have decided to 
find a value for $Z$ (distance from the Galactic plane), for which the space densities 
of thin and thick discs are almost equal, by performing Monte Carlo simulations with 
a wide range of parameters. For the Monte Carlo simulations in this study, wide 
ranges for the Solar space density and exponential scale height of 
the thick disc are adopted: 0\% $\leq n_{TK} \leq$ 15\% and 
500 $\leq H_{TK}\leq$ 1500 pc, respectively. The adopted exponential scale height range for 
the thin disc is 200 $\leq H_{TN}\leq$ 350 pc. Here, the subscripts TK and TN refer 
to the thick disc and the thin-disc components of the Galaxy, respectively. These parameter 
ranges were taken from \cite{Aketal07b}, who constructed a table of estimated parameters 
for the thin and thick discs from the literature. After 50000 trials for the 
Monte Carlo simulations, a histogram of the $Z$ values is obtained (Fig. 3). 
The mode value of 825 pc estimated from this histogram shows where 
the spatial densities of thin and thick discs are 
the most probably equal. This value is in agreement with those found in previous 
studies in which the deep sky surveys were used; cf. \cite{Ohjaetal1999} (0.79 kpc), 
\cite{Siegetal2002} (0.7-1 kpc), \cite{Karaalietal2004} (0.80-0.97 kpc), 
\cite{Biletal2006} (0.7-0.82 kpc). 

\begin{figure}[h]
\begin{center}
\includegraphics[scale=0.4, angle=0]{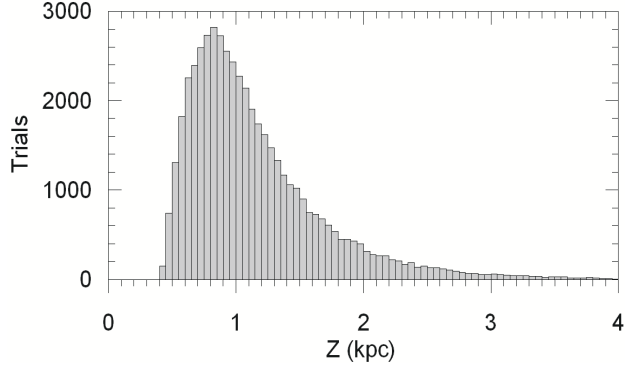}
\caption[] {\small Distribution of the vertical distances from the Galactic plane ($Z$), 
where space densities of the thin and thick discs are almost equal, obtained from the 
Monte Carlo simulations with 50000 trials for a wide range of Galactic model parameters.}
\end{center}
\end{figure}

Thus, in this study the CVs with $Z_{max}\leq$ 825 
pc are classified as thin-disc (TN) systems, while CVs with $Z_{max}>$ 825 pc 
are selected as thick disc or halo (TK-H) systems. Using this criterion, we 
have found that 17 of 216 CVs in the refined sample belong to the 
thick disc or halo population of the Galaxy. The rest of the sample is 
consisted of the thin-disc systems. The population types of the CVs in this 
study are indicated in Table 2. The columns of Table 2 are 
organized as name, Galactic coordinates ($l, b$), the corrected $U$, $V$ and $W$ 
components of the space velocity, maximum vertical distance to 
the Galactic plane ($Z_{max}$) and population type (TN or TK-H).

\begin{table*}
\setlength{\tabcolsep}{1.5pt} 
\begin{center}
\tiny{
\caption{The Galactic coordinates ($l, b$), corrected space velocity components ($U, V, W$), total 
space velocities ($S$),  maximum vertical distances from the Galactic plane of calculated orbits 
($Z_{max}$) and population types (Pop) of CVs in the preliminary sample. In the ``Pop'' column, 
TN and TK-H refers to thin disc and thick disc-or-halo CVs. Full 
table can be obtained electronically.}
\begin{tabular}{lccccccccccccc}
\hline
ID & Name  &      $l$     &     $b$      &      $U$      & $U_{err}$ &         $V$             &  $V_{err}$          &        $W$    & $W_{err}$           &  $S$          & $S_{err}$ &   $Z_{max}$  &  Pop \\
   &       & ($^{\circ}$) & ($^{\circ}$) & (km s$^{-1}$) & (km s$^{-1}$) & (km s$^{-1}$) & (km s$^{-1}$) & (km s$^{-1}$) & (km s$^{-1}$) & (km s$^{-1}$) & (km s$^{-1}$) & (kpc) &      \\
\hline
  1 &     WW Cet &     90.007 &    -71.743 &     -15.69 &       6.97 &      23.55 &       4.17 &     -14.27 &       6.01 &      31.69 &      10.10 &      0.312 &         TN \\
  2 &   V592 Cas &    118.603 &     -6.910 &       9.14 &       8.52 &      41.25 &      12.55 &       0.81 &       6.03 &      42.26 &      16.32 &      0.048 &         TN \\
  3 &   V709 Cas &    120.042 &     -3.455 &      21.46 &       3.86 &     -21.83 &       3.32 &       6.20 &       2.93 &      31.23 &       5.87 &      0.077 &         TN \\
  4 &     PX And &    116.992 &    -36.335 &      37.23 &      14.28 &      -2.11 &      19.35 &      -9.69 &      17.73 &      38.53 &      29.88 &      0.539 &         TN \\
  5 &    LTT 560 &    194.612 &    -88.105 &      19.94 &       2.69 &    -133.76 &      19.94 &     -29.18 &       0.73 &     138.35 &      20.13 &      0.395 &         TN \\
... &        ... &        ... &        ... &        ... &        ... &        ... &        ... &        ... &        ... &        ... &        ... &        ... &        ... \\
... &        ... &        ... &        ... &        ... &        ... &        ... &        ... &        ... &        ... &        ... &        ... &        ... &        ... \\
... &        ... &        ... &        ... &        ... &        ... &        ... &        ... &        ... &        ... &        ... &        ... &        ... &        ... \\
\hline     
\end{tabular}
}
\end{center}
\end{table*}

Representations of Galactic orbits, which are likely 
thick disc or halo CVs projected onto $X-Y$ and $X-Z$ planes, are shown 
in Fig. 4 and a list of these CVs are given in Table 4. 
$X$, $Y$ and $Z$ are heliocentric Galactic coordinates directed towards 
the Galactic Centre, Galactic rotation and the North Galactic Pole, 
respectively. The mean Galactocentric distances ($R_m$), and planar 
($e_p$) and vertical ($e_v$) orbital eccentricities of Galactic orbits 
calculated for thick disc or halo CVs are also given in Table 4. 
$R_m$ is defined as the arithmetic mean of the final perigalactic 
($R_p$) and apogalactic ($R_a$) distances of the Galactic 
orbit \citep{VidNin2009}, and $Z_{max}$ and $Z_{min}$ are the maximum 
and minimum distances, respectively, to the Galactic plane. 
$e_p$ and $e_v$ are defined as $e_{p}=(R_{a}-R_{p})/(R_{a}+R_{p})$ and 
$e_{v}=(|Z_{max}|+|Z_{min}|)/R_{m}$, respectively. The computed 
space velocity components of the magnetic and non-magnetic CVs in the 
refined sample are compared in the velocity space in  Fig. 5 according 
to the population types. As can be seen from Fig. 5, there is not 
a prominent difference between the velocity distributions of the magnetic 
and non-magnetic systems. In addition, thick-disc and halo CVs have 
higher $W$ velocities in the $V-W$ distribution (Fig. 5d) as compared 
to thin-disc stars.

\begin{figure*}[h]
\begin{center}
\includegraphics[scale=0.80, angle=0]{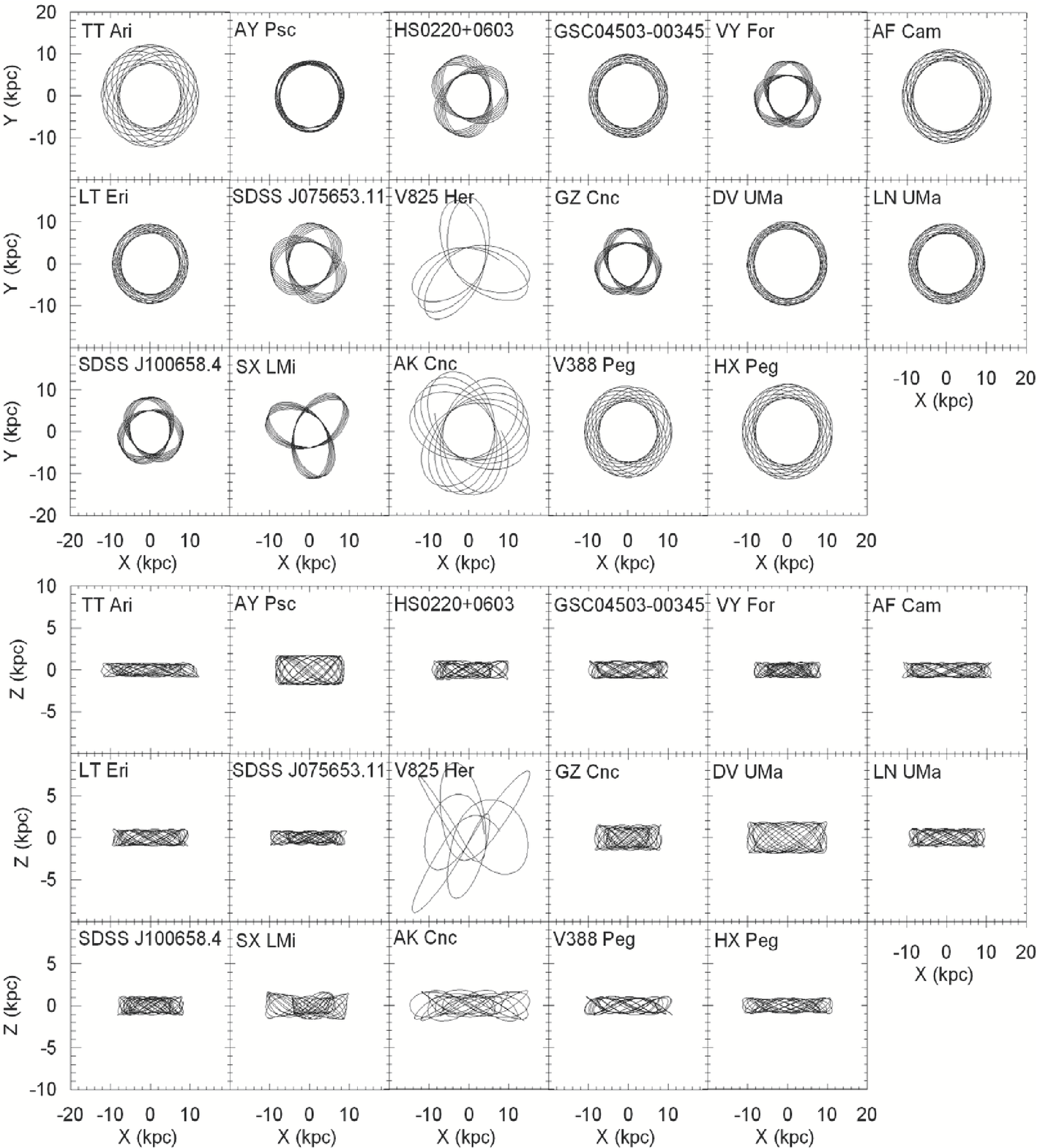}
\caption[] {\small Representations of the Galactic orbits computed for the likely 17 
thick disc or halo CVs projected onto $X-Y$ and $X-Z$ planes. Galactic orbits were 
calculated for an integration time of 3 Gyr.}
\end{center}
\end{figure*}

\begin{figure*}[h]
\begin{center}
\includegraphics[scale=0.8, angle=0]{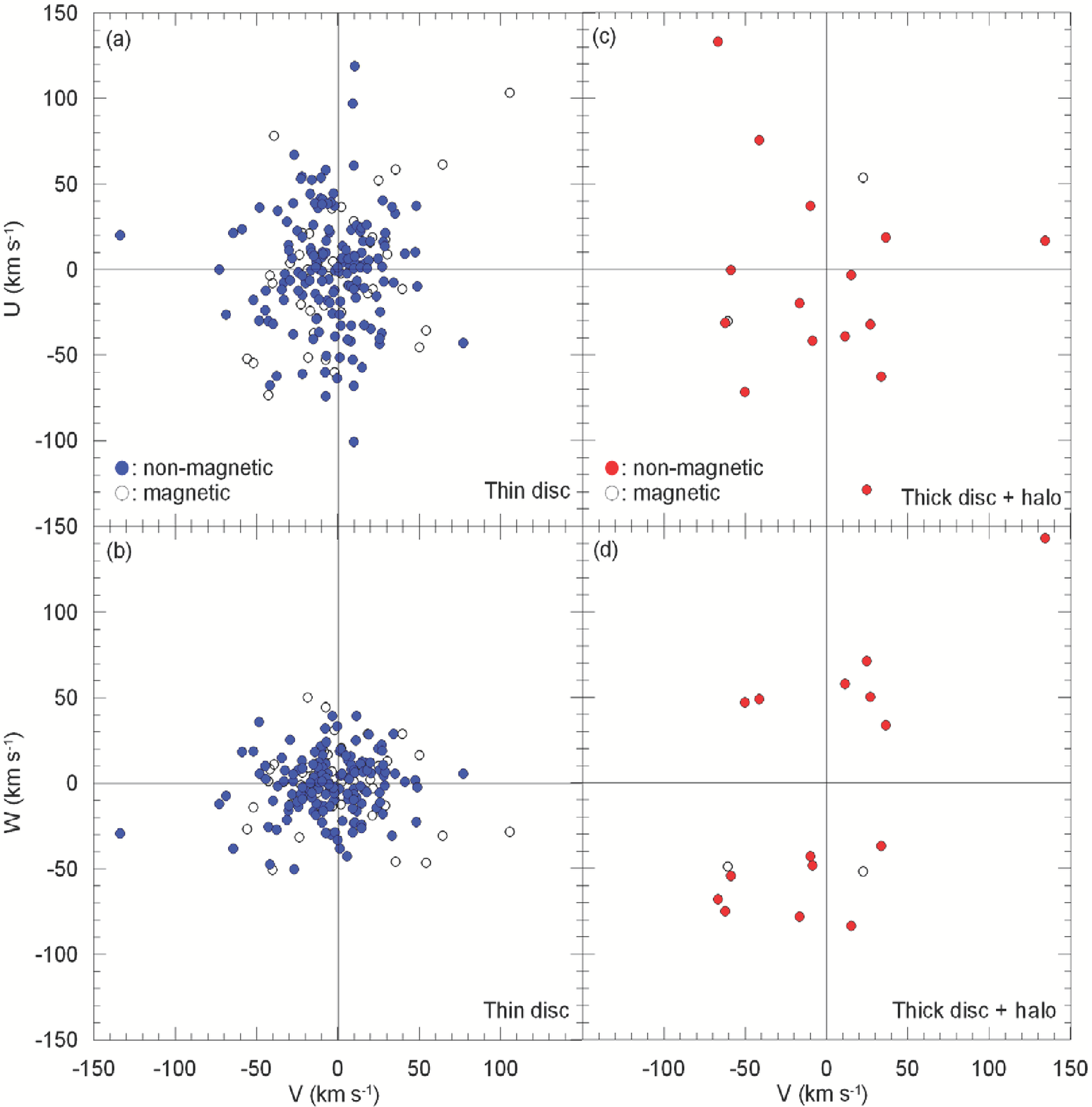}
\caption[] {\small The computed space velocity components of the magnetic 
and non-magnetic CVs in the refined sample according to the population types.}
\end{center}
\end{figure*}

\subsection{Velocity dispersions and kinematical ages} 

Kinematical age of a group of stars can be calculated from the velocity 
dispersion of the systems using the formulae given by \cite{Wielen1977}. For CVs 
in this study, we have used the age-space velocity dispersion relation improved 
by \cite{Cox00}: 

\begin{equation}
\sigma_{\nu}^{3}(\tau)=\sigma_{\nu,\tau=0}^{3}+\frac{3}{2}\alpha_{V}\delta_{2}T_{\delta}
\Biggl[\exp\Biggl(\frac{\tau}{T_{\delta}}\Biggl)-1\Biggr],
\end{equation}
where $\sigma_{\nu,\tau=0}$ is the velocity dispersion at zero age, which is 
usually taken as 10 km s$^{-1}$ \citep{Cox00}. $\alpha_{V}$ describes the rotation 
curve and it is taken approximately 2.95. $T_{\delta}$ is a time scale 
of $5\times10^{9}$ yr and  $\delta_{2}$ is a diffusion coefficient of 
$3.7\times10^{-6}$ (km~s$^{-1})^{3}$ yr. $\sigma_{\nu}(\tau)$ and $\tau$ are the 
total velocity dispersion and the kinematical age of the CV group in question, respectively. 
The connection between the total dispersion of space velocity vectors ($\sigma_{\nu}$) 
and the dispersion of the velocity components is described as 

\begin{equation}
\sigma_{\nu}^{2}=\sigma_{U}^{2}+\sigma_{V}^{2}+\sigma_{W}^{2}.
\end{equation}

After computing $\sigma_{\nu}^{2}$ from the dispersions of velocity components, 
the kinematical age $\tau$ for a group of systems can be easily computed by 
replacing this total dispersion in Eq. (4). Assuming an isotropic distribution 
for the systems, the $\gamma$ velocity dispersion is defined  
$\sigma_{\gamma}^{2}=(1/3)\sigma_{\nu}^{2}$ \citep{Wieetal92,vPaetal96}. 
The $\gamma$ velocity dispersions of CV groups according to different orbital 
period regimes and population types are computed to compare with their theoretical 
predictions and listed in Table 3.

\begin{landscape}  
\textwidth = 700 pt
\begin{table*}
\setlength{\tabcolsep}{2.1pt} 
\begin{center}
\scriptsize{
\caption{Mean space velocities, space velocity dispersions ($\sigma_{U}$, 
$\sigma_{V}$ and $\sigma_{W}$), total space velocity dispersions ($\sigma_{\nu}$), 
kinematical ages ($t$) and $\gamma$ velocity dispersions ($\sigma_{\gamma}$) of CV groups 
in the refined sample. ``All sample'' means all systems in the refined sample. M and 
nM denote magnetic (polars and intermediate polars) and non-magnetic systems, 
respectively. $N$ is the number of systems. TN represents thin-disc systems. The 
lower (2.15 h) and upper (3.18 h) borders for the period gap were adopted 
from \cite{Knig2006}. }
\begin{tabular}{lcccccccccc}
\hline
Parameter    & $N$   & $<U>$    & $<V>$    & $<W>$    & $\sigma_{U}$ & $\sigma_{V}$ & $\sigma_{W}$ & $\sigma_{\nu}$ & $t$           & $\sigma_{\gamma}$   \\
             &     &(km s$^{-1}$)   & (km s$^{-1}$)  & (km s$^{-1}$)  &      (km s$^{-1}$)   &     (km s$^{-1}$)    &     (km s$^{-1}$)    & (km s$^{-1}$)  & (Gyr)         & (km s$^{-1}$) \\
\hline
All sample   & 216 & -0.57$\pm$7.84 & -3.60$\pm$8.49 & -1.76$\pm$8.04 &      36.59$\pm$3.95  &     30.00$\pm$4.58   &    25.39$\pm$4.28    & 53.70$\pm$7.41 & 5.29$\pm$1.35 & 31.00$\pm$4.28   \\
\hline
$Z_{max} \leq 825$ pc  & 199 & 0.02$\pm$7.50 & -3.56$\pm$8.06 & -1.23$\pm$7.62 &      33.91$\pm$3.85  &     27.61$\pm$4.41   &    18.15$\pm$4.04    & 47.35$\pm$7.11 & 4.13$\pm$1.27 & 27.34$\pm$4.10   \\
$Z_{max} > 825$ pc     & 17  & -7.43$\pm$11.75 & -4.13$\pm$13.48 & -7.96$\pm$12.98 &      59.18$\pm$2.90  &     50.09$\pm$3.31   &    65.54$\pm$3.90    & 101.52$\pm$5.88 & 13.11$\pm$0.81 & 58.61$\pm$3.40   \\
\hline
Magnetic (M,TN)             & 41  &  0.47$\pm$7.70 &  0.11$\pm$8.17 & -1.24$\pm$8.18 &      38.53$\pm$4.52  &     33.10$\pm$4.48   &    22.60$\pm$4.21    & 55.60$\pm$7.63 & 5.64$\pm$1.39 & 32.10$\pm$4.41   \\
Non-Magnetic (nM,TN)        & 158 & -0.10$\pm$7.45 & -4.51$\pm$8.04 & -1.23$\pm$7.48 &      32.60$\pm$3.65  &     25.91$\pm$4.40   &    16.80$\pm$3.98    & 44.90$\pm$6.97 & 3.69$\pm$1.22 & 25.92$\pm$4.03   \\
\hline
$P < 2.15$ h (nM,TN) & 46  & -6.55$\pm$6.14 & -6.11$\pm$6.86 & 0.74$\pm$5.97  &      30.65$\pm$3.21  &     23.93$\pm$3.92   &    18.85$\pm$3.21    & 43.21$\pm$6.00 & 3.40$\pm$1.03 & 24.95$\pm$3.46   \\
$P > 3.18$ h (nM,TN) & 104 & 2.43$\pm$7.95  & -3.85$\pm$8.51 & -1.97$\pm$8.14 &      33.88$\pm$3.74  &     26.68$\pm$4.57   &    16.21$\pm$4.19    & 46.07$\pm$7.24 & 3.90$\pm$1.28 & 26.60$\pm$4.18   \\
\hline
$P < 2.15$ h (nM+M,TN)  & 56 & -2.74$\pm$6.07 & -3.03$\pm$6.84 & -0.36$\pm$6.07 &      33.63$\pm$3.25  &     28.42$\pm$4.06   &    20.07$\pm$3.45    & 48.39$\pm$6.24 & 4.32$\pm$1.12 & 27.94$\pm$3.60   \\
$P > 3.18$ h (nM+M,TN)  & 128 & 1.64$\pm$8.13 & -3.75$\pm$8.56 & -2.44$\pm$8.37 &      34.62$\pm$4.03  &     27.03$\pm$4.58   &    17.00$\pm$4.23    & 47.10$\pm$7.42 & 4.09$\pm$1.32 & 27.19$\pm$4.29   \\
\hline
$0.056 < P(d)\leq 0.080$ (nM+M,TN) & 48  & -0.92$\pm$5.73 & -3.47$\pm$6.55 & -1.53$\pm$5.93 &      33.93$\pm$2.98  &     28.91$\pm$3.93   &    20.52$\pm$3.56    & 49.07$\pm$6.08 & 4.44$\pm$1.10 & 28.33$\pm$3.51   \\
$0.080 < P(d)\leq 0.150$ (nM+M,TN) & 46  & -3.51$\pm$8.40 & -4.51$\pm$8.87 & 3.34$\pm$8.20 &      29.79$\pm$4.16  &     33.96$\pm$4.26   &    19.88$\pm$3.87    & 49.36$\pm$7.10 & 4.49$\pm$1.28 & 28.50$\pm$4.10   \\
$0.150 < P(d)\leq 0.210$ (nM+M,TN) & 50  &  1.46$\pm$8.06 & -5.50$\pm$8.12 & -5.59$\pm$7.93 &      35.52$\pm$3.66  &     25.15$\pm$4.58   &    17.06$\pm$3.92    & 46.75$\pm$7.05 & 4.02$\pm$1.25 & 26.99$\pm$4.07   \\
$0.210 < P(d)\leq 0.350$ (nM+M,TN) & 45  &  5.57$\pm$7.78 & -5.02$\pm$8.79 & -2.27$\pm$8.48 &      36.03$\pm$4.09  &     18.52$\pm$4.28   &    14.60$\pm$4.37    & 43.06$\pm$7.36 & 3.37$\pm$1.26 & 24.86$\pm$4.25   \\
$0.350 < P(d)\leq 0.500$ (nM+M,TN) & 10  &-11.46$\pm$7.89 & 16.65$\pm$8.10 & 5.71$\pm$7.68 &      27.97$\pm$3.05  &     26.59$\pm$5.19   &    7.55$\pm$3.57    & 39.32$\pm$6.98 & 2.74$\pm$1.13 & 22.70$\pm$4.03   \\
\hline
\end{tabular}
}
\end{center}
\end{table*}
\end{landscape}  

\section{Discussions}

In order to investigate the kinematical properties of CVs, we have collected 
a sample of the systems with proper motions and systemic velocities. 
Unfortunately not many CVs have trigonometric parallaxes except only 30 close 
ones. After estimating the rest of the distances using improved PLCs relations 
by \cite{Ozdonetal15} and \cite{Aketal07a}, their Galactic space velocity 
components were computed. A refined sample was constructed by eliminating 
CV’s with biggest errors $S_{err}>$ 30 km$s^{-1}$. In order to find out their 
Galactic population types, Galactic orbital parameters were computed. Additional 
sub-groups were formed in terms of their orbital periods.   

Kinematical properties of all sample and sub-groups are summarized in 
Table 3. The dispersions of space velocity components obtained from 
the refined sample as a whole are $\sigma_{U} = 36.59$ $\pm$ 3.95 km s$^{-1}$, 
$\sigma_{V} = 30.00$ $\pm$ 4.58 km s$^{-1}$, 
$\sigma_{W} = 25.39$ $\pm$ 4.28 km s$^{-1}$, indicating a mean kinematical 
age of 5.29 $\pm$ 1.35 Gyr. A systemic velocity ($\gamma$) dispersion of 
31.00 $\pm$ 4.28 km s$^{-1}$ is obtained by evaluating the total space 
velocity dispersion ($\sigma_{\nu}=53.70$ $\pm$ 7.41 km s$^{-1}$) of 
the refined sample. 

\subsection{Groups according to population types}

Galactic orbits of CVs in the refined sample shows that they are mostly located 
within the Galactic disc. Systems with vertical distances to the 
Galactic plane ($Z_{max}$) being larger than 825 pc are classified as the thick 
disc or halo CVs. We have found from the analysis of Galactic orbits that 
199 of 216 CVs in the refined sample are members of the thin-disc 
component of the Galaxy. The rest are likely to be the thick disc or halo 
systems. It must be noted that our sample is consisted of the systems in the 
Solar Neighbourhood. 

For further pinpointing the classification, we have calculated the planar 
($e_p$) and vertical ($e_v$) orbital eccentricities of the Galactic orbits 
in addition to the maximum ($Z_{max}$) and minimum ($Z_{min}$) vertical 
distances to the Galactic plane (Table 4). \cite{Bil12} had found from the 
distribution of the vertical orbital eccentricity of red clump stars that the stars 
with $e_{v}\leq 0.12$ and $0.12 < e_{v}\leq 0.25$ are the members of the thin and the 
thick disc populations of the Galaxy, respectively. Additionally, stars with 
$e_{v} > 0.25$ are halo objects. $Z_{max}$ values of seven systems in Table 4 are larger 
than 825 pc, while their vertical orbital eccentricities are $e_v\leq$ 0.12. 
Thus, following the classification scheme of \cite{Bil12}, we could conclude 
that these seven CVs can be in fact members of the 
old thin-disc population of the Galaxy. Nine of the 17 CVs in Table 4 are the 
thick-disc CVs. The total space velocity dispersion of the nine thick-disc CVs 
is found to be 93.87 $\pm$  5.13 km s$^{-1}$, which corresponds to a kinematical 
age of 12.0 $\pm$ 0.8 Gyr consistent with the age of the thick-disc component 
of the Galaxy \citep{Felt2009}. We have found one halo CV in our sample 
(V825 Her). The thick disc CVs in this study were not classified as the 
thick-disc members in \cite{Aketal13}. Disagreement between this study and 
\cite{Aketal13} possibly results from the new $\gamma$ velocities, new 
distance estimation method and different approximation to classify the systems. 

\begin{table*}
\setlength{\tabcolsep}{4pt} 
\begin{center}
\scriptsize{
\caption{Names, types, orbital periods ($P$), perigalactic ($R_p$) and apogalactic 
($R_a$) distances, the maximum ($Z_{max}$) and minimum ($Z_{min}$) vertical distances to the 
Galactic plane of calculated Galactic orbits for likely old thin disc, thick disc or halo 
stars in the refined sample. $e_p$ and $e_v$ denote planar and vertical orbital 
eccentricities of Galactic orbits for CVs. For the Type-1 and Type-2 columns, denotes are as 
described in Table 1. The most probable population types are indicated in the 
last column.}
\begin{tabular}{llcccccccccl}
\hline
ID & Name                      & Type-1  & Type-2  & $P$   & $R_p$ & $R_a$ &$Z_{max}$  & $Z_{min}$ & $e_p$ &  $e_v$    & Pop. Type      \\
   &                           &         &         &  (h)  & (kpc) & (kpc) &  (kpc)    &   (kpc)   &       &           &                 \\
\hline
1  & TT Tri                    &   NL    &  nM    & 3.3513 & 7.750  & 12.226 &    0.834    &   -0.833   &  0.22 & 0.08 & Old thin disc  \\
2  & AY Psc                    &   DN    &  nM    & 5.2157 & 7.491  &  8.705 &    1.770    &   -1.769   &  0.07 & 0.22 & Thick Disc     \\
3  & HS 0220+0603              &   NL    &  nM    & 3.5810 & 5.344  &  9.866 &    1.094    &   -1.090   &  0.30 & 0.14 & Thick Disc     \\
4  & GSC 04503-00345           &   NL    &  nM    & 3.8758 & 7.773  &  9.979 &    1.043    &   -1.043   &  0.12 & 0.12 & Old thin disc  \\
5  & VY For                    &   NL    &  M     & 3.8064 & 4.823  &  8.403 &    0.923    &   -0.922   &  0.27 & 0.14 & Thick Disc     \\
6  & AF Cam                    &   DN    &  nM    & 7.7779 & 8.527  & 11.218 &    0.918    &   -0.919   &  0.14 & 0.09 & Old thin disc  \\
7  & LT Eri                    &   DN    &  nM    & 4.0841 & 7.325  &  9.549 &    1.014    &   -1.014   &  0.13 & 0.12 & Old thin disc  \\
8  & SDSS J075653.11+085831.8  &   CV    &  nM    & 1.8960 & 5.224  &  9.774 &    0.871    &   -0.871   &  0.30 & 0.12 & Old thin disc  \\
9  & AK Cnc                    &   DN    &  nM    & 1.5624 & 6.612  & 15.251 &    1.850    &   -1.852   &  0.40 & 0.17 & Thick Disc     \\
10 & GZ Cnc                    &   DN    &  nM    & 2.1144 & 4.980  &  8.527 &    1.520    &   -1.519   &  0.26 & 0.23 & Thick Disc     \\
11 & DV UMa                    &   DN    &  nM    & 2.0605 & 8.359  & 10.136 &    1.892    &   -1.892   &  0.10 & 0.20 & Thick Disc     \\
12 & LN UMa                    &   NL    &  nM    & 3.4656 & 7.326  &  9.633 &    1.080    &   -1.079   &  0.14 & 0.13 & Thick Disc     \\
13 & SDSS J100658.40+233724.4  &   DN    &  nM    & 4.4619 & 5.023  &  8.344 &    1.092    &   -1.087   &  0.25 & 0.16 & Thick Disc     \\
14 & SX LMi                    &   DN    &  nM    & 1.6128 & 3.845  & 11.247 &    1.604    &   -1.603   &  0.49 & 0.21 & Thick Disc     \\
15 & V825 Her                  &   NL    &  nM    & 4.9440 & 7.839  & 32.136 &    8.930    &   -8.901   &  0.61 & 0.45 & Halo           \\
16 & V388 Peg                  &   NL    &  M     & 3.3751 & 7.278  & 11.010 &    1.091    &   -1.090   &  0.20 & 0.12 & Old thin disc  \\
17 & HX Peg                    &   DN    &  nM    & 4.8192 & 8.012  & 11.378 &    0.925    &   -0.923   &  0.17 & 0.10 & Old thin disc  \\
\hline     
\end{tabular}
}
\end{center}
\end{table*}

If seven likely old thin-disc CVs was included in the thin-disc group, 
number of the thin-disc CVs increase to 206. So, the space density of the thin-disc CVs 
increases to 95$\%$. Such a result would be in agreement with \cite{Aketal13} 
who claim 94$\%$ of Solar Neighbourhood CVs are thin-disc stars. This result 
is also in agreement with \cite{Peters08} who concluded that the kinematics 
of these systems are indicative of a thin-disc Galactic population. The space density 
of the thick-disc CVs in our sample ($\sim$5\%) shows that the refined CV sample 
in this study is complete for the Solar Neighbourhood, since space density of thick-disc 
CVs is in agreement with those derived for the field stars 
\citep{Robetal1996,Busetal1999,Biletal2006}. So, we conclude that statistical 
studies using the refined sample in our study give reliable results.

Kinematical properties of the thin disc ($Z_{max} \leq 825$ pc) and thick 
disc or halo ($Z_{max} > 825$ pc) CVs are listed in Table 3. Kinematical 
properties of these groups are very different from each other, as expected. Kinematical 
ages of the thin disc and thick disc or halo stars are 4.13 $\pm$ 1.27 and 
13.11 $\pm$ 0.81 Gyr, respectively. Although number of systems used in the estimation 
is small, kinematical ages of the CV groups from different Galactic population 
groups are consistent with the age of the Galactic components \citep{Wyse2013}.

\subsection{Magnetic and non-magnetic systems}

Kinematical ages of magnetic systems (polars and intermediate 
polars) could be different than non-magnetic systems, since the evolution of the 
magnetic systems  could be different than the evolution of non-magnetic CVs 
\citep{Wu1993,WebWick2002,Schetal07}. Thus, we have estimated the kinematical 
properties of the magnetic and the non-magnetic systems separately. Before doing 
this estimation, we have removed the thick disc and the halo CVs from the 
refined sample as the results can be biased by kinematics of these populations. 

Our results in Table 3 shows that non-magnetic thin-disc systems are younger than 
magnetic CVs: their kinematical ages are 3.69 $\pm$ 1.22 and 5.64 $\pm$ 1.39 
Gyr, respectively. Our results are roughly in agreement with those of \cite{Peters08} 
who estimated mean kinematical ages of $\geq$4 and $\geq$5 Gyr for non-magnetic and 
magnetic systems, respectively. However, \cite{Aketal10} found an age of 7.68 $\pm$ 
1.44 Gyr for magnetic systems since they did not remove thick disc and halo CVs 
from their sample which biased the age estimates.  

\subsection{Groups according to orbital periods}

Table 3 shows that the kinematical ages of non-magnetic thin-disc CVs below 
($P < 2.15$ h) and above ($P > 3.18$ h) the orbital period gap, which 
are 3.40 $\pm$ 1.03 and 3.90 $\pm$ 1.28 Gyr, respectively. When we consider 
magnetic and non-magnetic thin-disc systems together, their ages are estimated as 
4.32 $\pm$ 1.12 and 4.09 $\pm$ 1.32 Gyr for the systems below and above the gap, 
respectively.  

Standard evolution theory predicts that the CVs above the period gap 
must be younger than the systems below the gap. If this is true, the kinematical 
properties of the non-magnetic thin-disc CVs must obey this rule in the absence of 
the bias from the thick disc or the halo systems. However, our results are not in 
agreement with this prediction. It is clear from Table 3 that there is not 
a considerable age difference between the thin-disc non-magnetic systems below 
and above the orbital period gap. Even if we include magnetic systems in the age 
calculation, which are considerably older than non-magnetic CVs, kinematical ages 
for the systems below and above the gap remain almost equal. These results are in 
agreement with \cite{Aketal10} and \cite{vPaetal96}. 

It is predicted from the age structure of a Galactic CV population obtained 
by applying standard formation and evolution models \citep{KS96} that CVs 
above the orbital period gap must have an average age of 1~Gyr, while the mean 
age of systems below the gap should be 3-4~Gyr \citep[see also][]{RB86}. This 
age difference is mainly due to the time spent evolving from the post-CE phase 
into the contact phase \citep{Kol01}. Although the age 3.40 Gyr derived for the systems 
below the gap is in agreement with the theoretical prediction, we can not find 
a 2-3~Gyr age difference for the systems above and below the gap. 

In order to investigate the age differences between the various period regimes, we 
have divided the refined sample into smaller subsamples according to almost the same 
number of CVs at different period ranges. The total space velocity dispersions 
and corresponding kinematical ages for CVs in these period ranges are summarized 
in Table 3. $U-V$ diagrams for the systems in these period ranges are 
shown in Fig. 6. Unlike the age-period relation presented in \cite{Aketal10}, 
Table 3 shows that the kinematical ages decrease towards longer orbital periods, 
as expected from the standard evolution theory, with a decreasing rate 
of  $dP/dt=-1.62(\pm 0.15)\times 10^{-5}$ sec yr$^{-1}$. It seems that the 
systems near the upper border of the gap affect the mean age estimate of systems 
above the gap and increase the mean value, when all systems above the gap are 
taken into account as a CV group. Thus, we can conclude that the space velocity 
dispersion, and so the age, decreases as the orbital period increases. 

\begin{figure*}[h]
\begin{center}
\includegraphics[scale=0.8, angle=0]{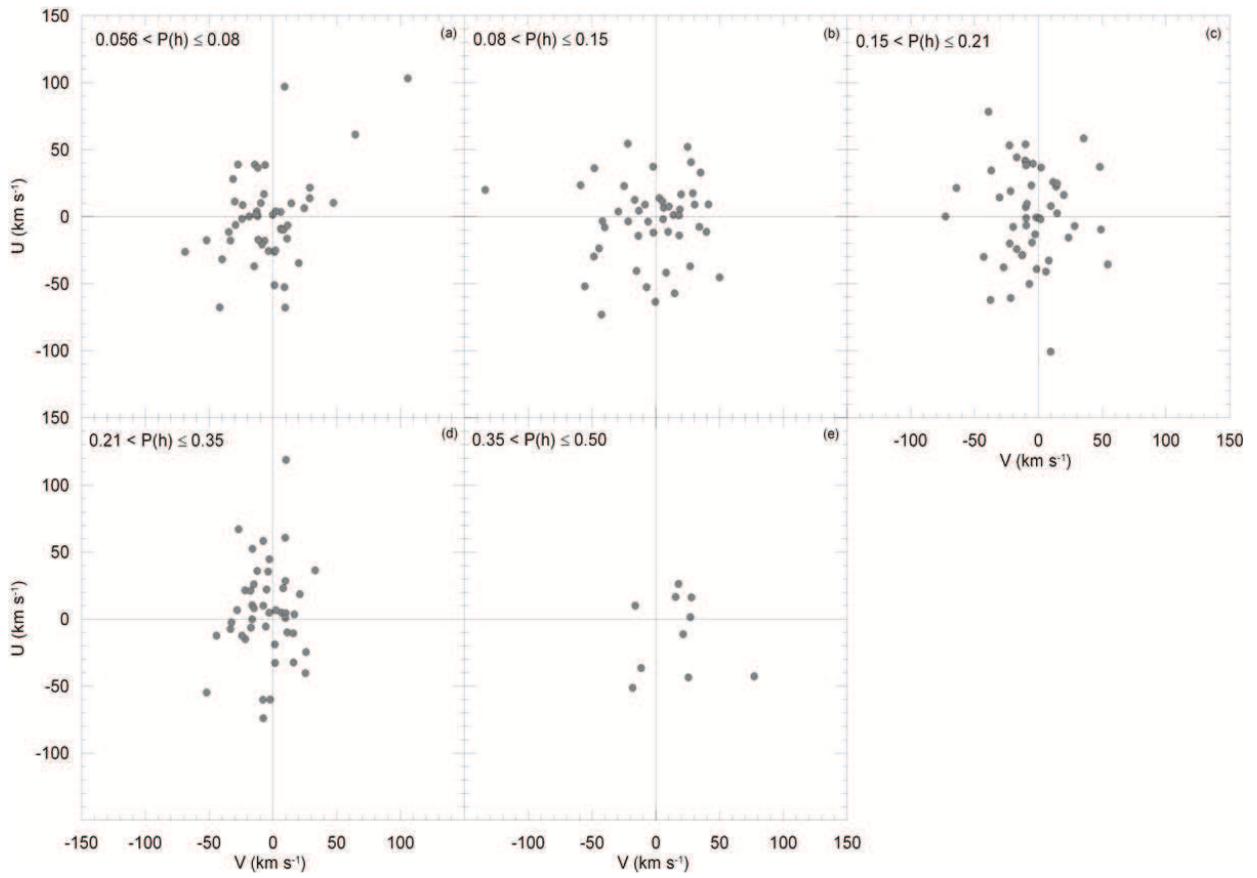}
\caption[] {\small $U-V$ diagrams of CVs in the refined sample for different 
orbital period ranges.}
\end{center}
\end{figure*}

\subsection{The $\gamma$ velocity dispersion of CVs}

The $\gamma$ velocity dispersions of CV sub-groups are useful tools to 
test predictions of the standard formation and evolution theory. The 
$\gamma$ velocity dispersion of all systems in the refined sample is estimated as 
$\sigma_\gamma=31.00\pm 4.28$ km s$^{-1}$. This value is not significantly 
different than $\sigma_\gamma=30\pm 4$ km s$^{-1}$ found by \cite{Aketal10}. 
\cite{KS96} predicted that the $\gamma$ velocity dispersions for the systems 
above and below the orbital period gap are $\sigma_{\gamma} \simeq 15$ and 
$\sigma_{\gamma} \simeq 30$ km~s$^{-1}$, respectively. Prediction of \cite{Kol01} 
states that the $\gamma$ velocity dispersions of CVs should be 
$\sigma_{\gamma} \simeq 27$ and $\sigma_{\gamma} \simeq 32$ km~s$^{-1}$ for 
the systems above and below the gap, respectively, if magnetic braking does 
not operate in the detached phase. These theoretical predictions suggest that there 
must be considerable difference for the $\gamma$ velocity dispersions of CVs 
below and above the orbital period gap. In order to compare with these theoretical 
predictions, we have calculated the $\gamma$ velocity dispersions for all thin-disc 
systems below and above the period gap, and found $\sigma_\gamma=27.94\pm 3.60$ 
km s$^{-1}$ and $\sigma_\gamma=27.19\pm 4.29$ km s$^{-1}$, respectively. Although 
the theoretically predicted dispersion for CVs below the gap is roughly in agreement 
with observations, it is clear that CVs below and above the gap have $\gamma$ 
velocity dispersions that are the same, within the errors. A substantial amount of 
difference between the $\gamma$ velocity dispersions of the systems below and above 
the gap can not be obtained, as well, even if we use only non-magnetic thin-disc CVs 
(Table 3): $\sigma_\gamma=24.95\pm 3.46$ and $\sigma_\gamma=26.60\pm 4.18$ km s$^{-1}$ 
for below and above the gap, respectively. Note that using only $\sigma_{U}$ and 
$\sigma_{V}$ for the calculation of total dispersion does not change this similarity 
between the $\gamma$ velocity dispersions. 

\section{Conclusions}

By analysing available kinematical data of CVs, we have concluded that there is not 
considerable kinematical difference between the systems below and above the orbital 
period gap. This result is not in agreement with the standard theory of the CV 
evolution and results of \cite{Aketal10}. Thus, we can not confirm the prediction of 
\cite{KS96} who predicted 2-3 Gyr mean age difference between the CVs below and above 
the period gap. Smaller age difference implies similar angular momentum loss time 
scales for systems with low-mass and high-mass secondaries \citep{Kol01}. 
However, it must be noted that kinematical age of CV groups slightly increases with 
decreasing orbital period, a result in agreement with the standard theory of the 
CV evolution. 

Observational $\gamma$ velocity dispersion of CVs below the period gap is roughly in 
agreement with the predictions of the standard theory CV evolution \citep{KS96,Kol01}. 
However, a substantial amount of difference for $\gamma$ velocity dispersions of the 
systems below and above the period gap could not be obtained from the observations. 

By calculating Galactic orbital parameters of CVs in the sample, we found that 17 of them 
are likely members of old thin disc, thick disc or halo components of the Galaxy. 
Orbital eccentricities and maximum vertical distances to the Galactic plane show that 
only one of them is a halo CV (V825 Her). From the population analysis based on a 
pure dynamical approximation, we have concluded that CVs are very consistent with the 
thin-disc population of the Galaxy. 

\section{Acknowledgments}

Part of this work was supported by the Research Fund of the University of Istanbul, 
Project Numbers: 27839, 39170 and 39742. This work has been supported in part by 
the Scientific and Technological Research Council (T\"UB\. ITAK) grand numbers: 111T224 
and 212T091. This research has made use of the SIMBAD database, operated at 
CDS, Strasbourg, France. This research has made use of NASA's Astrophysics Data System.

\end{document}